\documentstyle[aaspp4,12pt]{article}

\def\etal{{\it et al.}}

\def\ROSAT{{\it ROSAT}}
\def\AXAF{{\it AXAF}}
\def\EUVE{{\it EUVE}}
\def\mic100{\mbox{100$\mu$m}}
\def\F100{\mbox{F$_{\nu}$(100$\mu$m)}}

\def\NHxo{{\rm N_{\rm H,x}}}
\def\NHIo{{\rm N_{\rm HI}}}
\def\NH21{{\rm N_{\rm H,21cm}}}

\def\acm2{cm$^{-2}$}

\def\H2{H$_{2}$}

\def\pg{$\pm$\phn}
\def\ph{\phn$\pm$\phn}

\def\Msun{M$_{\odot}$}
\def\ayr{y$^{-1}$}

\begin{document}
\newcommand{\ARAA}[2]{\it Ann. Rev. Astron. Astrophys., \rm#1, #2.}
\newcommand{\ApJ}[2]{\it Ap. J., \rm#1, #2.}
\newcommand{\ApJL}[2]{\it Ap. J. Lett., \rm#1, #2.}
\newcommand{\ApJSS}[2]{\it Ap. J. Supp. Ser., \rm#1, #2.}
\newcommand{\AandA}[2]{\it Astron. Astrophys., \rm#1, #2.}
\newcommand{\AJ}[2]{\it Astron. J., \rm#1, #2.}
\newcommand{\BAAS}[2]{\it Bull. Amer. Ast. Soc., \rm#1, #2.}
\newcommand{\ASP}[2]{\it Astron. Soc. Pac. Conf. Ser., \rm#1, #2.}
\newcommand{\JCP}[2]{\it J. Comp. Phys., \rm#1, #2.}
\newcommand{\MNRAS}[2]{\it M. N. R. A. S., \rm#1, #2.}
\newcommand{\N}[2]{\it Nature, \rm#1, #2.}
\newcommand{\PASJ}[2]{\it Publ. Astron. Soc. Jap., \rm#1, #2.}
\newcommand{\RPP}[2]{\it Rep. Prog. Phys., \rm#1, #2.}
\newcommand{\tenup}[1]{\times 10^{#1}}

\title{On the Extreme Ultraviolet Emission from Galaxy Clusters}
\author{John S.\ Arabadjis and Joel N.\ Bregman}
\affil{Dept.\ of Astronomy, University of Michigan \\
Ann Arbor, MI 48109-1090 \\
jsa@astro.lsa.umich.edu, jbregman@umich.edu}

\begin{abstract}

An extremely soft X-ray excess throughout galaxy clusters has been claimed
as a new feature of these sytems, with important physical implications.  We
have reexamined this feature in the five clusters for which it has been
discussed, using the most recent X-ray absorption cross sections, X-ray data
processing techniques, and a consistent set of HI data.  For the Virgo
cluster, we find that the spectrum can be fit with a single-temperature
thermal plasma and with an X-ray absorption column that is not significantly
different than the Galactic HI column.  The result for Abell 1367, Abell 1656
(Coma), Abell 1795, and Abell 2199 is similar in that the difference between
the X-ray absorption column and the Galactic HI column is less than 3$\sigma$
for He/H = 0.09, and for He/H = 0.10 only one cluster location leads to a
Galactic HI column more than 3$\sigma$ above the X-ray absorption column
(Coma, with one location with a 3.6$\sigma$ difference).  We conclude that
there is no strong evidence for the extremely soft X-ray excess in galaxy
clusters.

\end{abstract}

\section{Introduction}

One of the surprises in the studies of galaxy clusters is that they were
detected by the \EUVE\ (Lieu \etal\ \markcite{lmblhs2}1996c), an instrument
whose primarily goals were the studies of stars and gas in the local
neighborhood.  The \EUVE\ has both imaging and spectroscopic capabilities
that operate in the spectral range 70-760 \AA\ (177-16.3 eV; Haisch, Bowyer,
and Malina \markcite{hbm}1993).  For high latitude sight lines of low
Galactic $\NHIo$ ($1.0\tenup{20}$ \acm2), the optical depth $\tau=2.8$ at
130 eV and $\tau=1.16$ at 180 eV, so 6-30\% of the emission will be
unabsorbed, permitting the detection of bright soft X-ray sources.  Soft
X-ray emission also can be detected with the \ROSAT\ PSPC, which has an
energy response extending below 150 eV and has significantly more collecting
area than the \EUVE\ at energies of significant transmission of X-rays.
At 0.13 keV (the peak sensitivity of the \EUVE\ Lexan/boron detector)
the effective area is 28 cm$^2$, compared with 40 cm$^2$ for the \ROSAT\ PSPC.
At 0.155 keV, the \EUVE\ effective area has dropped to less than cm$^2$,
whereas for the PSPC it is greater than 100 cm$^2$.

In their study of the Virgo cluster, centered on M87, Lieu \etal\
\markcite{lmblhs1}(1996b) were unable to successfully fit the data with a
single-temperature spectral model at the ambient cluster temperature (about
2 keV) and with the X-ray absorbing column fixed at the Galactic $\NHIo$.
The failure to fit the data was due to a large excess of soft residuals,
indicating that there was an additional soft component.  A two-temperature
model led to an acceptable fit when the soft component had a temperature of
about 0.05 keV, implying that large amounts of gas at $5\tenup{5}$K are
present.  An exciting consequence of this observation is that the cooling
rate of the gas, if in a steady-state, would be at least 340 \Msun\ \ayr,
at least a factor of 30 greater than the value determined from the standard
cooling flow picture of 10 \Msun\ \ayr\ (Fabian, Nulsen, and Canizares
\markcite{fnc}1984).

The presence of such a large amount of cooling gas has a variety of
consequences, such as the a mass of low temperature component that was
comparable to the virial mass of the cluster (Fabian \markcite{fab}1996;
Sarazin and Lieu \markcite{sl}1998).  Also, there were observable
consequences, such as emission from the O VI species, which was not detected
in the same clusters in which the soft component was present (Dixon, Hurwitz,
and Ferguson \markcite{dhf}1996).  Faced with these difficulties, another
suggestion was proposed for this emission: the soft X-ray component was due
to inverse Compton radiation of the cosmic microwave background by low energy
cosmic ray electrons (Sarazin and Lieu 1998).

Since the original study of the Virgo cluster, several other clusters have
been studied by the same group: Abell 1656 (Coma), Abell 1795, Abell 2199,
and Abell 1367 (Lieu \etal\ \markcite{lmbblmh}1996a,b,c; Mittaz, Lieu, and
Lockman \markcite{mll}1998; Table~\ref{tab:list}).  In each case, the excess
X-ray emission is detected at approximately the same energy, about 0.15-0.25
keV.  One might expect a thermal feature to occur at the same energy in
different clusters, but it seems surprising to us that a nonthermal component
would always appear at the same energy.

Here we take a different approach to examining the phenomenon of the excess
soft emission.  Since the soft emission becomes less prominent, and may
disappear if lower Galactic absorption columns were possible, we ask whether
it is feasible to achieve a spectral fit without the soft component and for
absorption columns consistent with the Galactic $\NHIo$.  Although we do not
disagree with the fitting method employed by Lieu and collaborators, there
has recently been a downward revision of the X-ray cross section at these
energies, due to improved cross sections for \mbox{He I}.  We examine whether
this change to the X-ray cross section permits an acceptable spectral fit
without a soft component.

\section{Data Processing and Analysis
\label{datproc}}

One of the central issues in the measurement of the soft component in galaxy
clusters is the correction for the absorption of X-rays by Galactic gas.  A
very important issue for this absorption, which has only been rectified
recently, is the value for the cross section of He (Fig.~ref{f1}).  At
energies in the 150-250 eV range, where most of the absorption occurs, He
accounts for about 71\% of the cross section, with hydrogen providing the
remainder.  The commonly used cross section of Ba{\l}uci{\'{n}}ska-Church and
McCammon (\markcite{bcm}1993) adopted a cross section for He that is based
upon data from Marr and West (\markcite{mw}1976).  These are 18-19\%
larger than the recent determination of Yan, Sadeghpour, and Dalgarno
(\markcite{ysd}1998), which is similar to the determinations of Samson \etal\
(\markcite{shyh}1994), Bizeau and Wuilleumier (\markcite{bw}1995), and
Morrison and McCammon (\markcite{mm}1983, which is in turn based upon data
from Henke \etal\ \markcite{hltsf}1982).  By using the improved cross section
for He, the total cross section at 150 eV decreases by 13\%, thus causing a
rise in the expected soft continuum in a single temperature fit (for fixed
$\NHIo$).

Using the new absorption cross sections, we have addressed the issue of
whether the X-ray spectra from the clusters showing soft component can be fit
without employing a soft component.  In this case, we let the value for the
Galactic absorption column be a parameter that is fit, rather than fixing it.
If a successful fit is discovered, as determined from an acceptable $\chi^2$,
we examine whether the fit value for the Galactic absorption column is
consistent or inconsistent with the Galactic $\NHIo$ value.
 
The X-ray spectral fitting is performed on our data as discussed by Arabadjis
and Bregman (\markcite{ab}1998), whereby the \ROSAT\ PSPC data are corrected
for gain fluctuations across the image plane (through PCPICOR) and periods of
time with high backgrounds are removed, which removes only a few percent of
the data at most.  Two concentric adjacent annuli, with point sources removed,
were used to produce a pair of spectrally well-behaved X-ray sources for each
cluster (Table~\ref{tab:source}).  (In A2199 and A1795 the annuli are chosen
specifically to avoid the known cooling flow regions.)  The exception to this
is Coma (Abell 1656), where we sought to avoid the galaxies near the center of
the cluster.  In that case we chose two circular regions with radii of
$3^{\prime}$ near the center but avoiding the member galaxies.  Background
spectra were generally taken from annuli with widths between 2-4$^{\prime}$
and radii between 15 and $20^{\prime}$, again with point sources removed.

The temperatures and redshift of each cluster were taken from White, Jones and
Forman (\markcite{wjf}1997), and the metallicity assumed for each cluster was
0.3.  It should be noted that other recent temperature determinations (e.g.
White \etal\ \markcite{wdhh}1994; Donnelly \etal\ \markcite{dmfjdcg}1998) are
within a few tenths of a keV of our adopted values, the effect of which is
minimal upon the derived columns.  This leaves two free parameters in each
fit, the intervening column and the spectral normalization.  The exception to
this is Virgo, where we fit for the temperature along with the column and
normalization, and set the abundance to 0.34 (Hwang \etal\
\markcite{hmlmfm}1997).  The resulting temperatures, $1.60 \pm 0.06$ and
$1.67 \pm 0.11$ keV, are consistent with the temperature profiles derived for
Virgo by Nulsen and B\"{o}hringer \markcite{nb}(1995) and D'Acri, De Grandi,
and Molendi \markcite{dadgm}(1998).  The derived columns are not particularly
sensitive to changes of the order of $\delta T \sim 0.2$ K and
$\delta z \sim 0.2$, changing by less than 5\%.

Spectral fitting is performed using XSPEC version 10 (see, e.g., Arnaud
\markcite{arn}1996), with the important change that we have put in the new
cross sections for \mbox{He I} of Yan, Sadeghpour, and Dalgarno (1998) into
the X-ray absorption routine (VPHABS).  Unlike the work discussed by Lieu
\etal\ (1996a,b,c), we use the \ROSAT\ data over the energy range 0.14-2.4 keV
but we avoid the three softest channels, for which the calibration may be
unreliable (Briel, Burkert, and Pfeffermann \markcite{bbp}1989; Snowden
\etal\ \markcite{stgy}1995).
 
Successful single component fits were obtained for each of the objects
(Table~\ref{tab:fits}) for helium abundances of He/H=0.09 and 0.10.  This
range brackets most recent abundance determinations (Osterbrock, Tran, and
Veilleux \markcite{otv}1992; Baldwin \etal\ \markcite{bfmccps}1991), although
Dupuis \etal\ (\markcite{dvbpt}1994) find a somewhat lower value based upon
\EUVE\ observations of DA stars.  Our model fits show a tight correlation
between the He abundance and the column density:  a reduction of 10\% in He/H
increases the derived column by $6.5 \pm 0.5$\%.  As we discuss later, the
effect is to bring the X-ray columns closer to their corresponding Galactic
HI columns.

Spectral fits with He/H=0.10 are shown with their residuals in
Figure~\ref{f2}.  In region 1 of A1656 and A1795 and region 2 of A2199 the
residuals show a small systematic modulation near 0.25 keV.  This is due to a
small error in the \ROSAT\ calibration matrices which results in a slight
offset in the peak of the response function.  Because the offset is in the
negative direction, it does not change our conclusion:  if it is not due to
calibration errors, it implies a either soft X-ray {\it deficit}, or a
somewhat greater absorption column.  As discussed above, in the case of Virgo
the temperature was an additional fit parameter in the spectral modelling, and
so we show confidence contours in $\NHxo$ and T in Figure~\ref{f3}.

\section{Discussion and Conclusions}

The study of the Virgo cluster was the work that established the studies of
soft excess emission in clusters and motivated further work so we begin our
discussion here.  Our analysis leads to differences between the X-ray
absorption column and the 21 cm HI column that are no greater than 1.3$\sigma$
and the residuals are symmetric about zero (Table~\ref{tab:fits};
Figure~\ref{f2}).  The difference between our result and that of Lieu \etal\
(1996b) is due primarily to the lower cross section that we used and
secondarily to a lower 21 cm column density derived by Hartmann and Burton
(\markcite{hb}1997).  The difference between the 21 cm HI column of Hartmann
and Burton (1997) and of Lieu \etal\ (1996b) is within the uncertainties of
the two different techniques used and is probably caused by small calibration
differences, as previously discussed in Arabadjis and Bregman (1998).  When we
use the 21 cm columns and cross sections used by Lieu \etal\ (1996b) we find
excess soft emission as well, which nearly disappears when the intervening
column in the model is allowed to be fit.  Figure~\ref{f4}a is a
monotemperature fit to the Virgo data which uses the Lieu \etal\ (1996c) value
$\NHxo = 1.8\tenup{20}$ \acm2 and fits for the cluster temperature, abundance,
and metallicity.  There does indeed appear to be excess emission from 0.14-0.30
keV, and the reduced chi-squared value of the fit, $\chi^2_r$, is an
unacceptable 1.37.  The Lieu study remedied this by introducing a second
(lower) temperature component which lowered $\chi^2_r$ to a marginally
acceptable 1.3.  Our fit (Figure~\ref{f4}b), which uses the Yan \etal\ (1998)
\mbox{He I} cross section and allows $\NHxo$ to vary, eliminates the soft
excess as well, with $\chi^2_r=1.12$.  We note that the X-ray column densities
that we obtained are similar to those obtained by Nulsen and B\"{o}hringer
(1995), who used the absorption cross sections of Morrison and McCammon
(1983), which are very similar to the Yan, Sadeghpour, and Dalgarno (1998)
cross sections that we employed (see Figure~\ref{f1}).  In summary, our X-ray
absorption column densities, which are similar to some other studies, are not
significantly different than the 21 cm HI column density, leading us to
conclude that a soft X-ray excess at energies below 0.3 keV is not a required
feature of the X-ray spectrum of the Virgo cluster.

For the other four clusters, the X-ray absorption column is generally within
2.5$\sigma$ of the 21 cm column when He/H = 0.10, and within 1.8$\sigma$ of
the 21 cm column for H/He = 0.09 (Table~\ref{tab:fits}; Figure~\ref{f5}). 
In Abell 2199, the X-ray column exceeds the 21 cm column, while for the three
other clusters, it is lower than the 21 cm column.  However, in only one
location is the X-ray absorption column more than 3$\sigma$ different than
the 21 cm column, position 2 in the Coma cluster when using the larger He/H
value of 0.10 (a 3.6$\sigma$ difference).  We find this rather weak evidence
for concluding that the X-ray absorption column is lower than the 21 cm
column.
 
As discussed previously (Arabadjis and Bregman 1998), these findings imply
that the ionized layer responsible for the pulsar dispersion measure must be
very highly ionized (at least 50\% of the He in the form of \mbox{He III}),
since it cannot contribute significantly to the X-ray absorption column.
However, the ionized column is consistent with that associated with a hot
Galactic halo, which has been confirmed as a feature of the Galaxy by two
independent groups (Pietz \etal\ \markcite{pkkbhm}1998; Snowden \etal\
\markcite{seffp}1998).

Future work will be able to reduce the uncertainties for several of these
measurements.  The upcoming X-ray telescope \AXAF\ should have an excellent
calibration and it will have greatly superior spectral resolution compared to
\ROSAT, so the accuracy of the fit and the resulting determination of the
X-ray absorbing column should be more accurate.  Also, more accurate 21 cm
measurements of the HI column will be possible with the Green Bank Telescope,
which should become operational within the next year.  However, a major source
of uncertainty is the He/H ratio in the ISM, and unless future observations
can help to decide how to handle dust corrections to photionization with
improved accuracy, this uncertainty will persist.
 
The authors would like to thank J. Irwin, M. Sulkanen, S. Snowden, F. Lockman,
D. Hartmann, and C. Sarazin for their comments and suggestions that assited us
in this investigation.  We would like to acknowledge financial support from
NASA grant NAG5-3247.

\vspace{20pt}

\begin{deluxetable}{rrrcl}
\tablewidth{250pt}
\tablecaption{The five clusters in the sample.
\label{tab:list}}
\tablehead{
\colhead{cluster}        &
\colhead{{\it l}$^{II}$} &
\colhead{{\it b}$^{II}$} &
\colhead{T, keV}\tablenotemark{a} &
\colhead{z}\tablenotemark{a}
}
\startdata
  A1367  & 234.80 &  +73.03 & 3.5     & 0.0214 \nl
  A1656  &  58.16 &  +88.01 & 8.0     & 0.0231 \nl
  A1795  &  33.81 &  +77.18 & 5.1     & 0.0621 \nl
  A2199  &  62.93 &  +43.69 & 4.7     & 0.0299 \nl
  Virgo  & 283.78 &  +74.49 & 1.6,1.7 & 0.0037 \nl
\enddata
\tablenotetext{a}{The temperature T and redshift z are taken from White,
Jones and Forman (1997) and references therein, except for Virgo, whose
temperature in the two source regions are determined from a $\chi^2$ fit.}
\end{deluxetable}

\clearpage
\begin{deluxetable}{cccccc}
\tablewidth{450pt}
\tablecaption{Observation and data analysis parameters.
\label{tab:source}}
\tablehead{
\colhead{cluster}  &
\colhead{$t_{\rm int}$\tablenotemark{a}, ks}  &
\colhead{annulus\tablenotemark{b} (1), $^{\prime}$}  &
\colhead{photons (1)}  &
\colhead{annulus\tablenotemark{b} (2), $^{\prime}$}  &
\colhead{photons (2)}
}
\startdata
 A1367 & 18.1  & 2-8\phn  & {\phn}8777  &  {\phn}8-14    &      12070 \nl
 A1656 & 20.4  & 0-3\tablenotemark{c}\phn
   & {\phn}7508  &  {\phn}0-3\tablenotemark{c}\phn
   & {\phn}4739 \nl
 A1795 & 35.1  & 3-6\phn  &      42567  &  {\phn}6-9\phn &      25613 \nl
 A2199 & 41.1  & 11-12     &     11190  &  {\phn}0-3\tablenotemark{d}\phn
   &      11673 \nl
 Virgo & 10.1  & 9-12     &      33775  &       12-15    &      31659 \nl
\enddata
\tablenotetext{a}{$t_{\rm int}$ is the integration time for the observation.}
\tablenotetext{b}{Two concentric source annuli (or disks in the case of Abell
1656) were constructed for each cluster, delineated by the inner and outer
radius.}
\tablenotetext{c}{The source regions in Abell 1656 consist of two circular
regions near the emission center with radii of $3^{\prime}$.  These were
chosen in order to avoid galaxies near the cluster center.}
\tablenotetext{d}{In order to avoid bright point sources the second source
region in Abell 2199 is a circular region near the emission center with
a radius of $3^{\prime}$.}
\end{deluxetable}

\clearpage
\begin{deluxetable}{rcllclcc}
\tablewidth{480pt}
\tablecaption{Column densities toward each cluster in the sample (in units of
$10^{20}$ \acm2).
\label{tab:fits}}
\tablehead{
\colhead{cluster}                      &
\colhead{He/H}                         &
\colhead{$\NHxo(1)$\tablenotemark{a}}  &
\colhead{$\NHxo(2)$\tablenotemark{a}}  &
\colhead{$\NH21$\tablenotemark{b}}     &
\colhead{N$_{\rm e}$\tablenotemark{c}} &
\colhead{$\frac{\Delta \rm{N}}{\sigma}(1)$\tablenotemark{d}} &
\colhead{$\frac{\Delta \rm{N}}{\sigma}(2)$\tablenotemark{d}}
}
\startdata
A1367 & 0.10& 1.87  \ph 0.16  & \phn 1.65  \ph 0.19  & 2.20 \ph 0.110 & 0.531 &
$-1.70$ & $-2.51$ \nl
      & 0.09& 2.00  \ph 0.17  & \phn 1.77  \ph 0.21  &                & &
$-0.99$ & $-1.81$ \nl
A1656 & 0.10& 0.781 \pg 0.050 & \phn 0.597 \pg 0.071 & 0.90 \ph 0.045 & 0.509 &
$-1.77$ & $-3.61$ \nl
      & 0.09& 0.836 \pg 0.053 & \phn 0.639 \pg 0.076 &                & &
$-0.92$ & $-2.96$ \nl
A1795 & 0.10& 0.909 \pg 0.026 & \phn 0.909 \pg 0.017 & 1.04 \ph 0.052 & 0.525 &
$-2.25$ & $-2.40$ \nl
      & 0.09& 0.964 \pg 0.028 & \phn 0.963 \pg 0.019 &                & &
$-1.29$ & $-1.33$ \nl
A2199 & 0.10& 0.830 \pg 0.035 & \phn 0.877 \pg 0.047 & 0.81 \ph 0.041 & 0.744 &
$+0.37$ & $+1.07$ \nl
      & 0.09& 0.889 \pg 0.037 & \phn 0.939 \pg 0.051 &                & &
$+1.43$ & $+1.97$ \nl
Virgo & 0.10& 1.73  \ph 0.055 & \phn 1.59  \ph 0.079 & 1.72 \ph 0.086 & 0.531 &
$+0.10$ & $-1.11$ \nl
      & 0.09& 1.86  \ph 0.060 & \phn 1.71  \ph 0.085 &                & &
$+1.34$ & $-0.08$ \nl
\enddata
\tablenotetext{a}{$\NHxo$ and $1\sigma$ errors toward two different source
regions in each cluster.}
\tablenotetext{b}{$\NH21$ from Hartmann and Burton (1997).}
\tablenotetext{c}{N$_{\rm e}$ calculated using the model of Taylor and Cordes
(\markcite{tc}1993).}
\tablenotetext{d}{$\Delta$N$/\sigma = (\NHxo - \NH21)/
(\sigma_{\rm Xray}^2+\sigma_{\rm 21cm}^2)^{1/2}$.}
\end{deluxetable}

\clearpage
\plotone{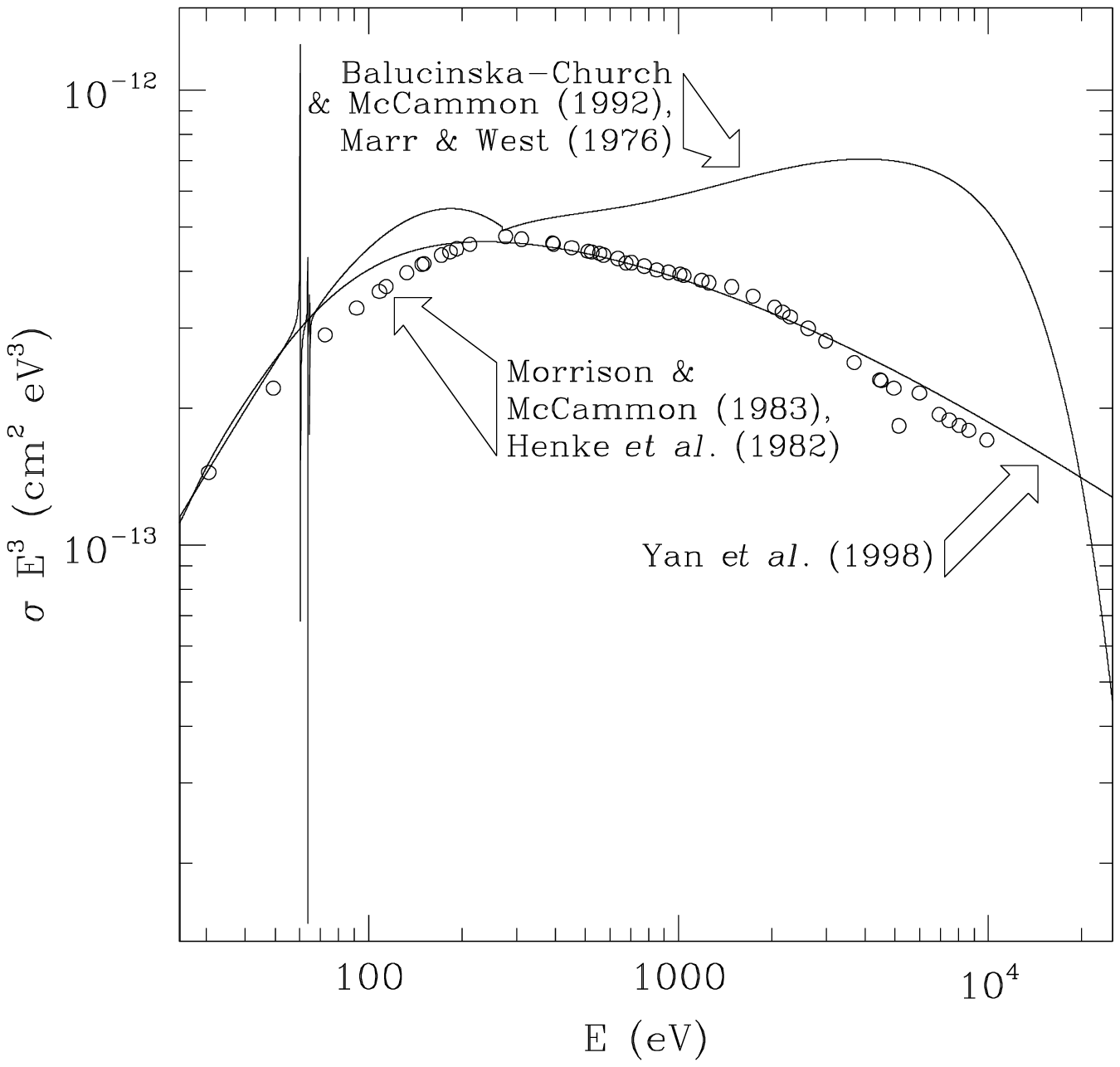}

\figcaption{A comparison of neutral helium cross sections.  The autoionization
features near \mbox{E $\sim 60$ eV} are shown only for the
Ba{\l}uci{\'{n}}ska-Church and McCammon (1992) cross sections.  These are not
relevant to our models, however, since the fitting range of this study is
0.14-2.0 keV.
\label{f1}}

\clearpage
\plotone{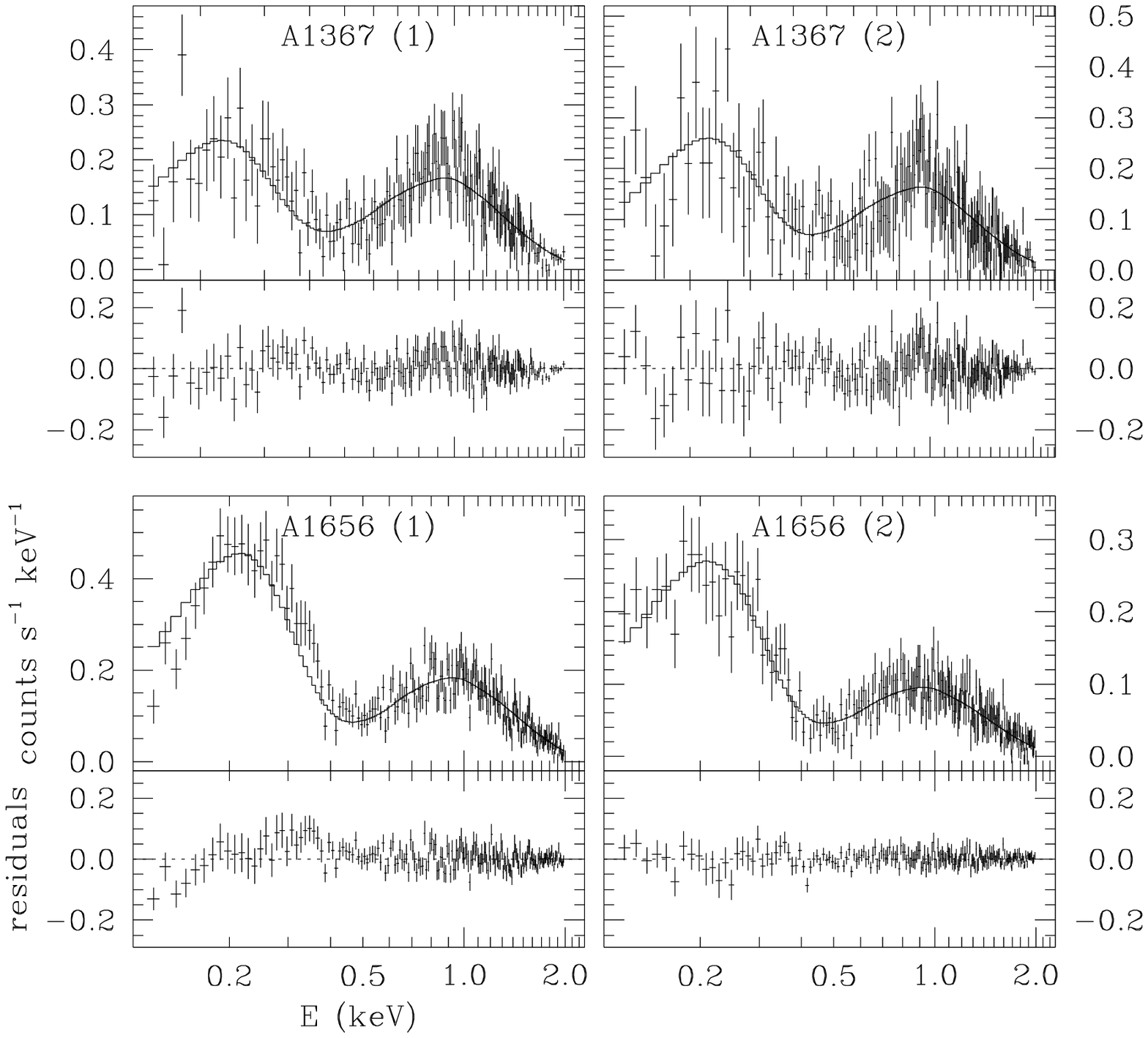}

\clearpage
\plotone{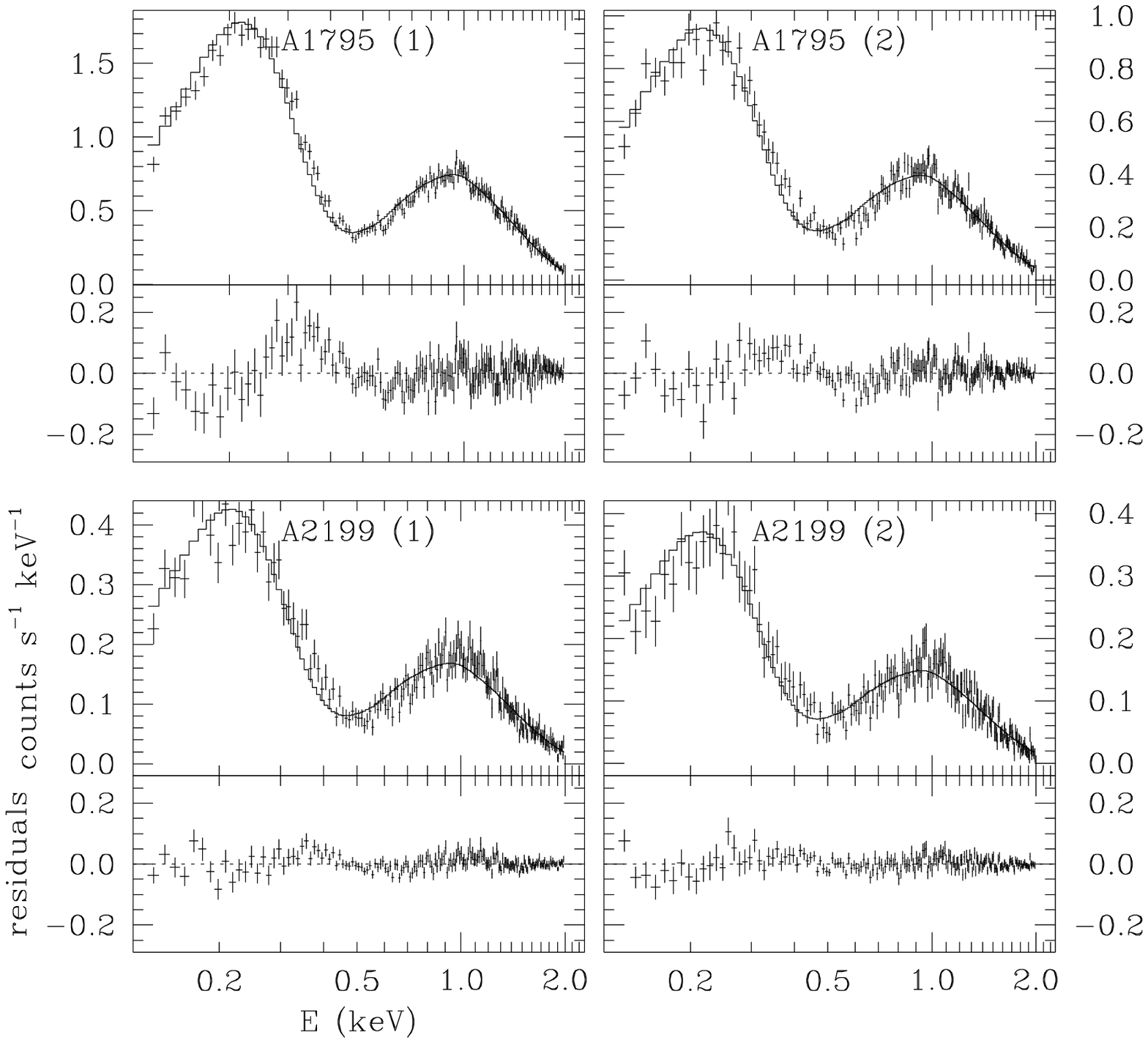}

\clearpage
\plotone{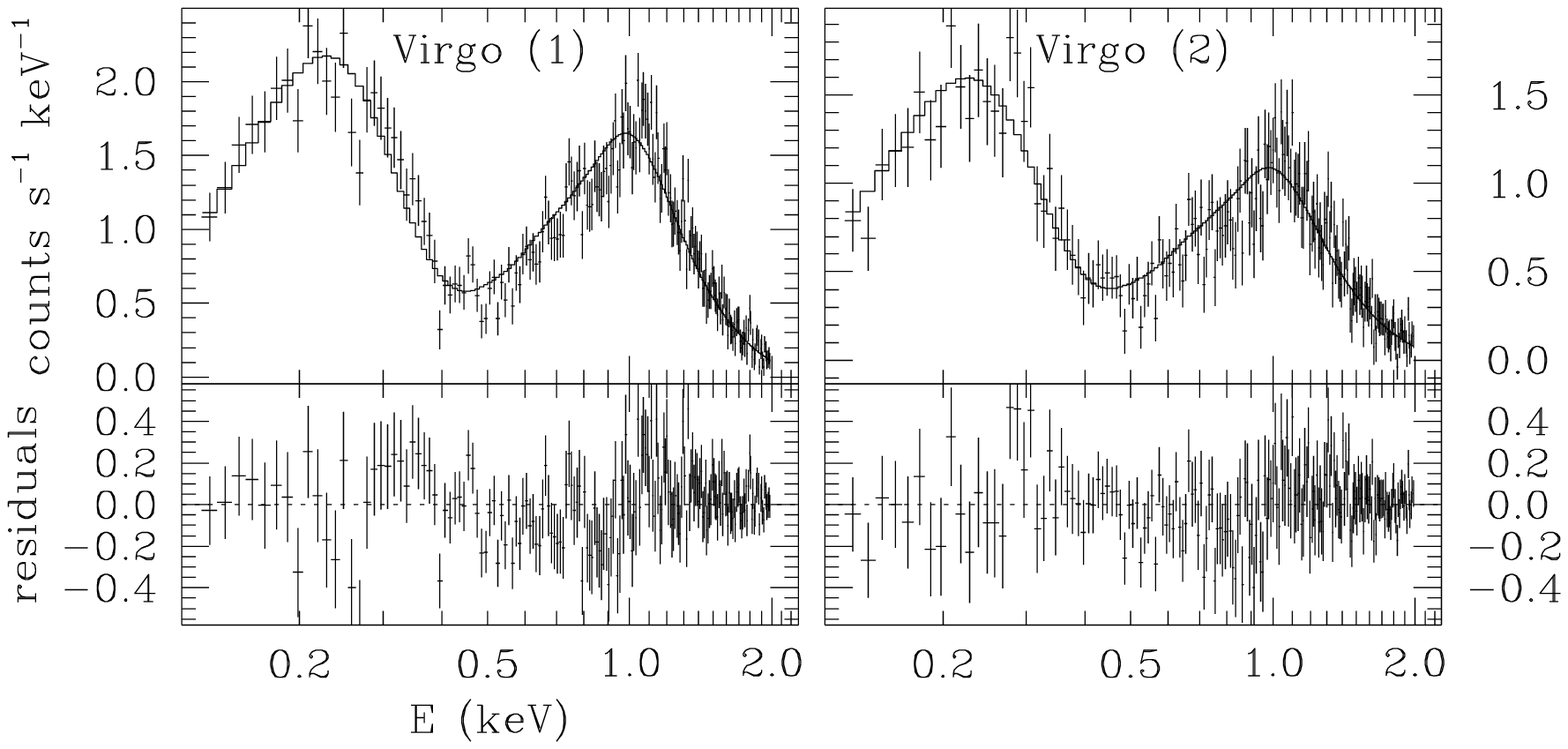}

\figcaption{Spectral fits and residuals for the five clusters for a helium
abundance of He/H=0.10.
\label{f2}}

\clearpage
\plotone{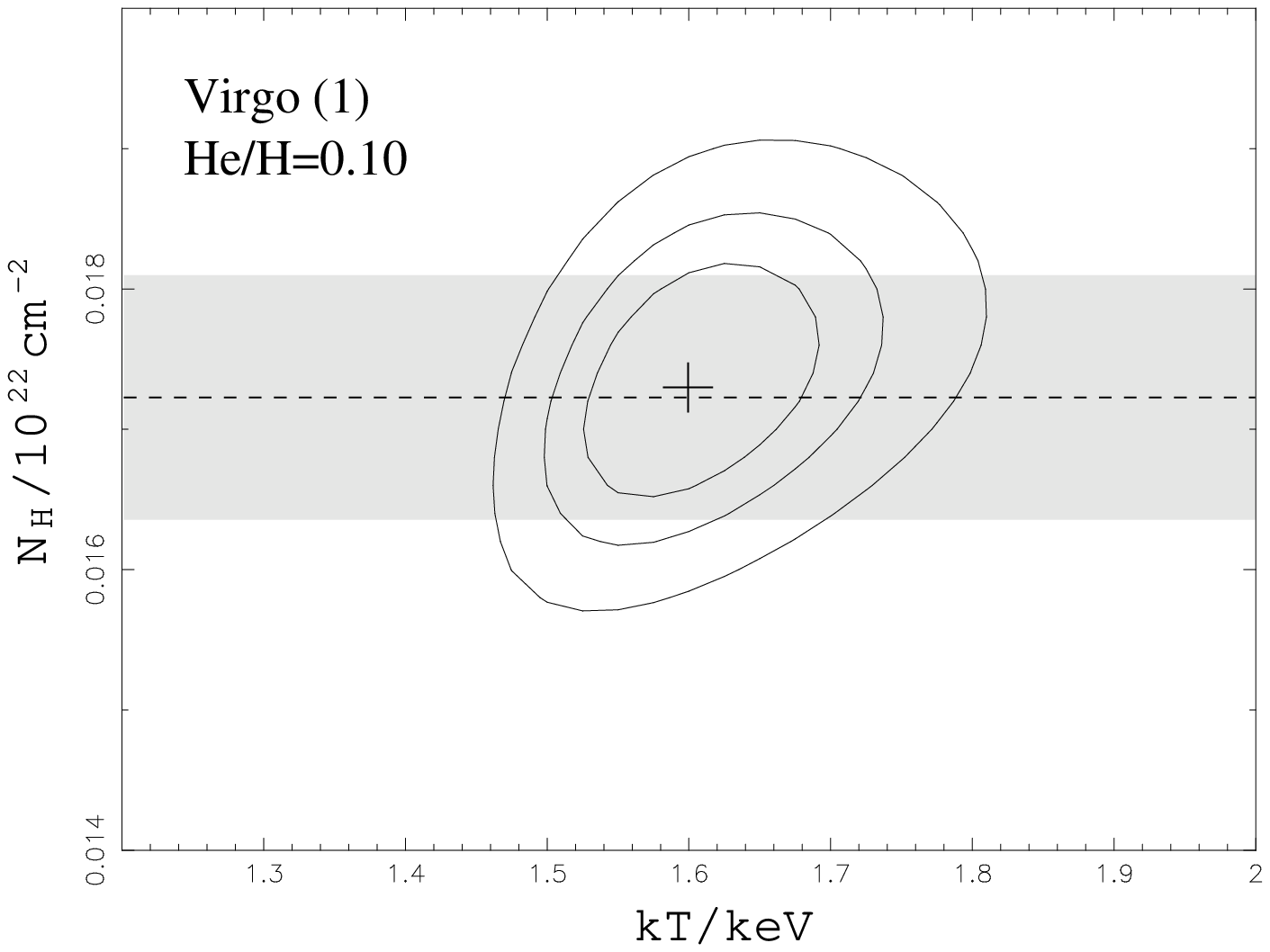}

\clearpage
\plotone{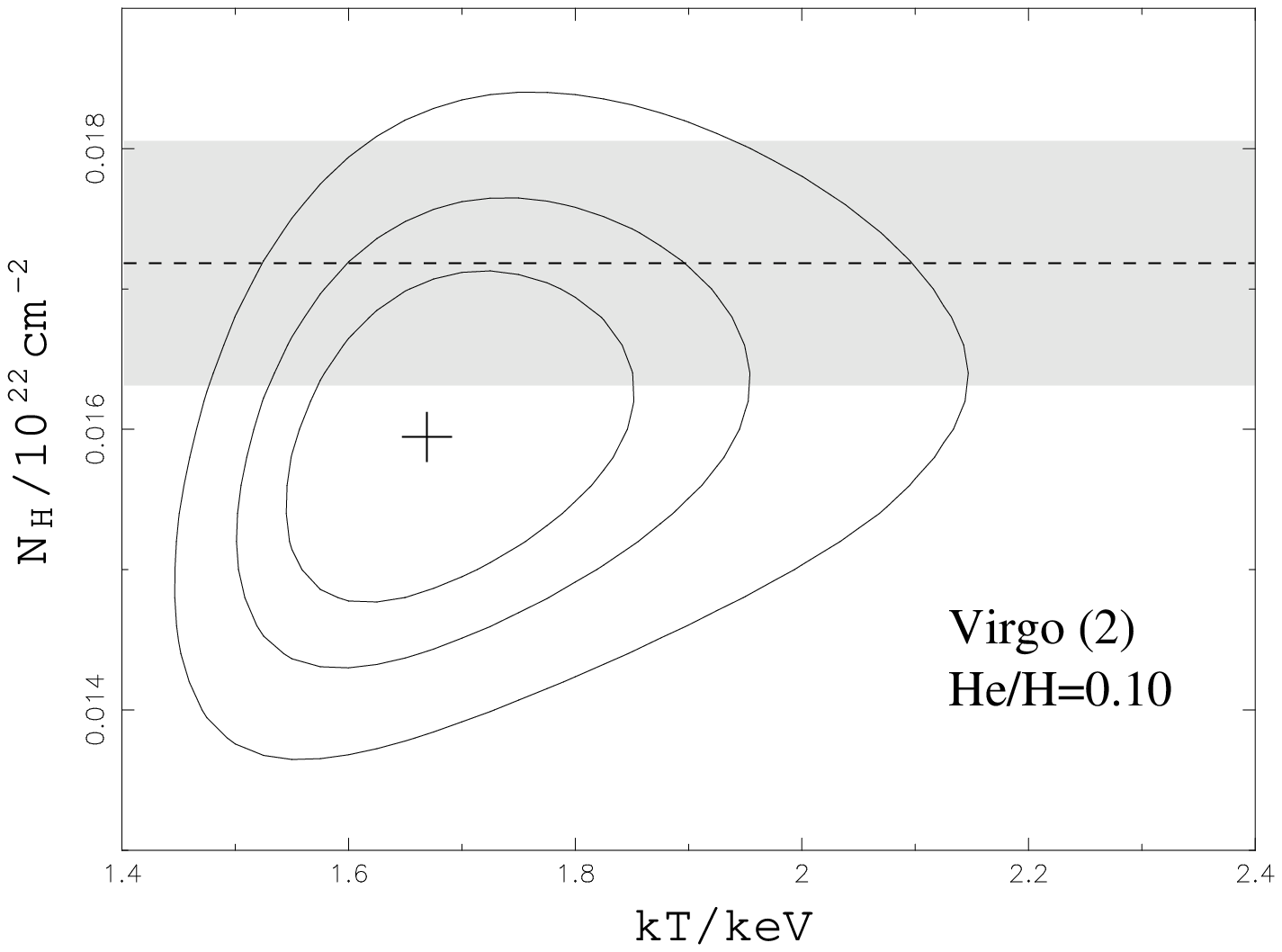}

\clearpage
\plotone{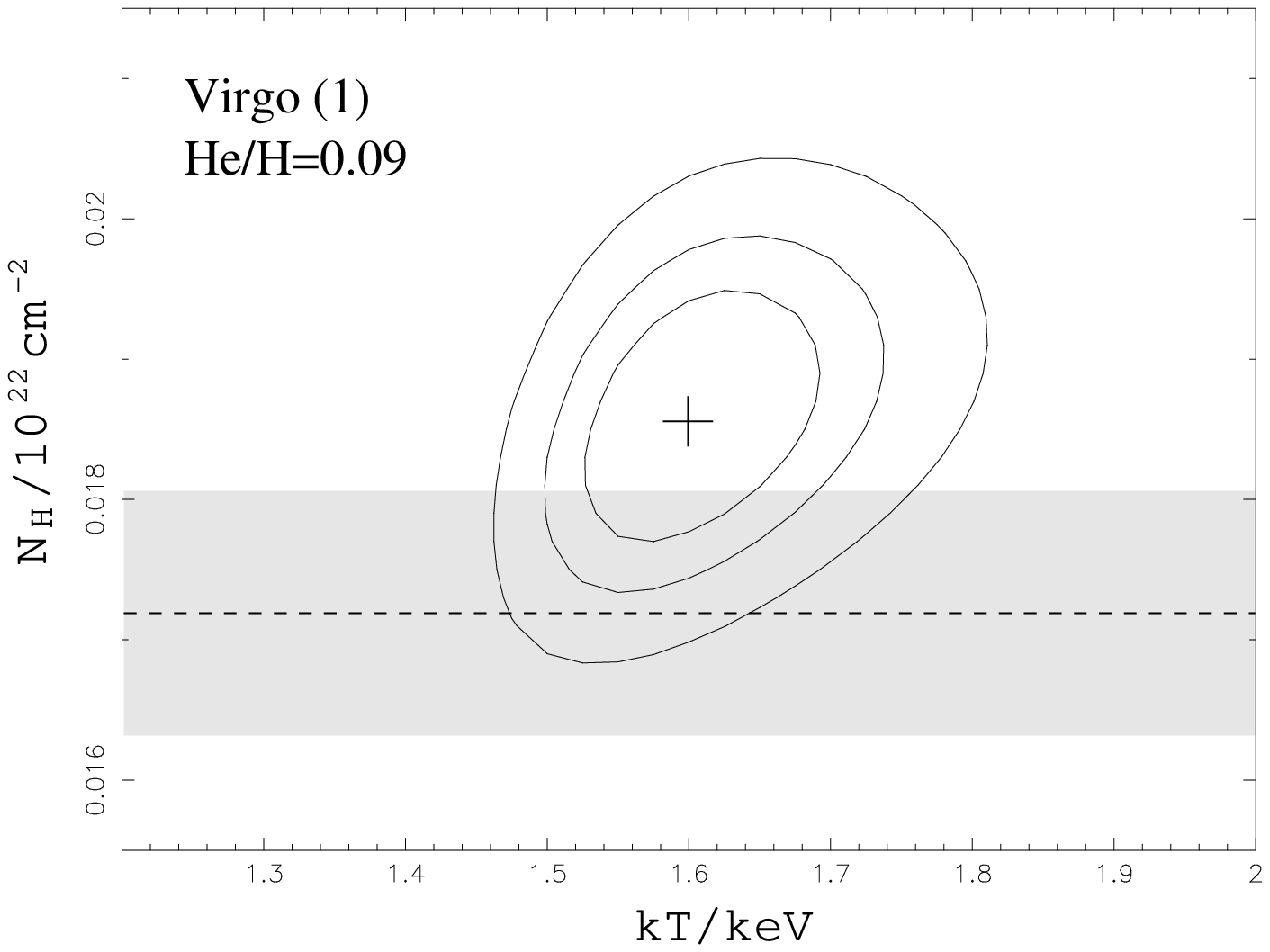}

\clearpage
\plotone{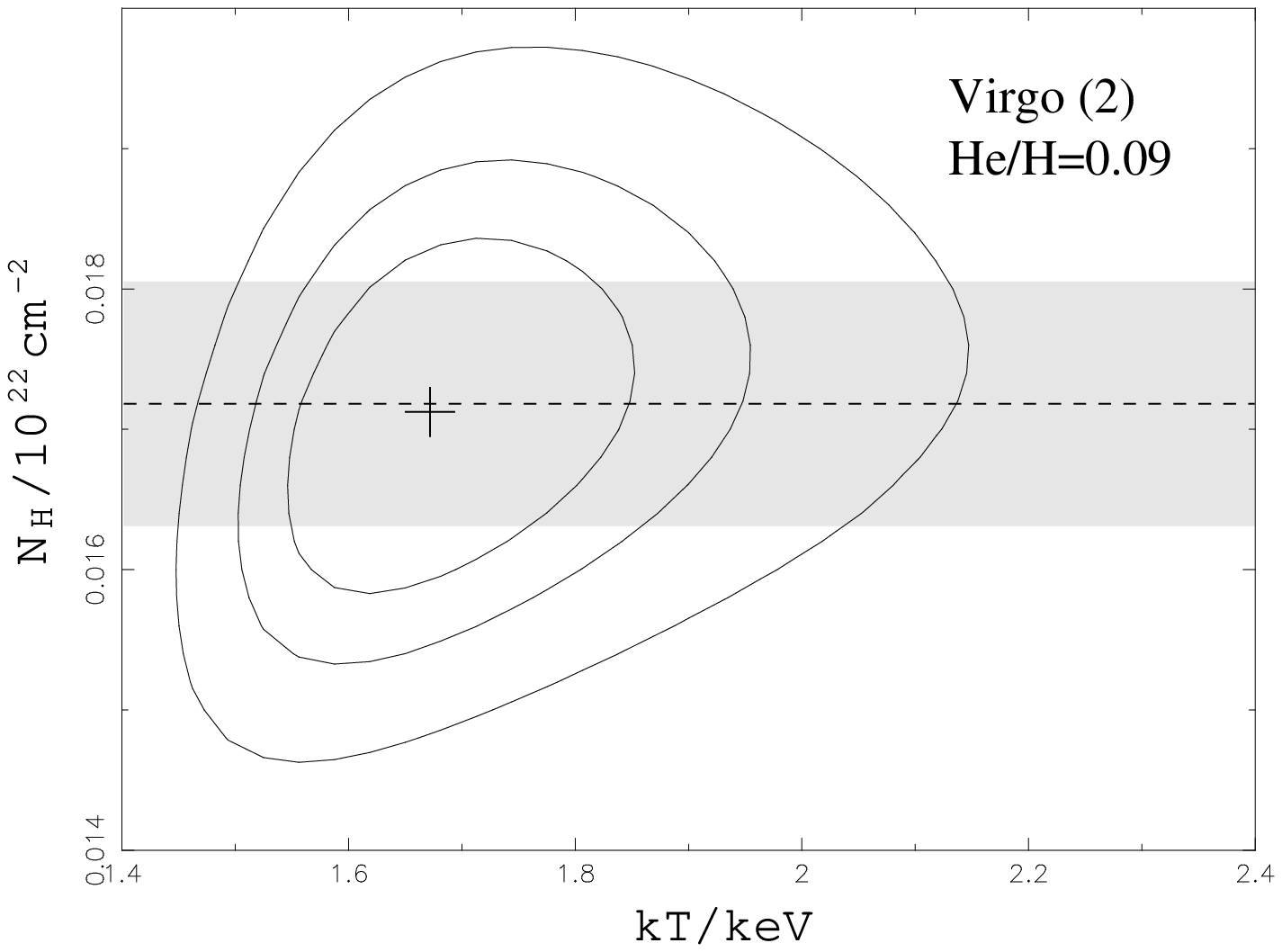}

\figcaption{$\Delta \chi^2$ contours in parameters T and $\NHxo$ for the Virgo
spectral fits.  Moving outward, the countours have values of 2.3, 4.6, and
9.2, corresponding to confidence limits of 68\%, 90\%, and 99\%.  The crosses
indicate T and $\NHxo$ of best-fit model.  The 21 cm column and uncertainty
(Hartmann and Burton 1997) is shown by the dashed lines and shaded regions,
respectively.
\label{f3}}

\clearpage
\plotone{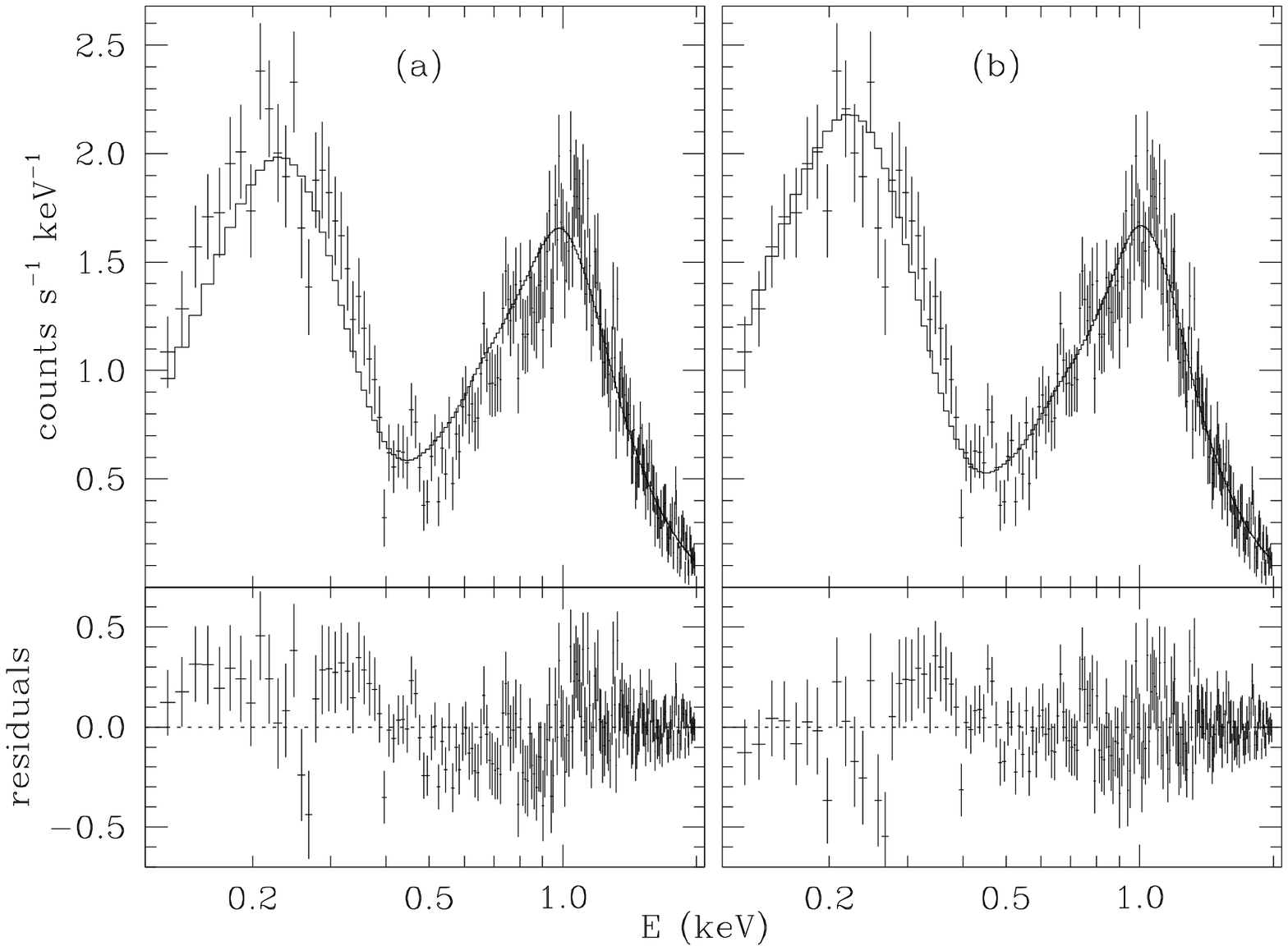}

\figcaption{A monotemperature fit to Virgo using (a) $\NHxo=1.8\tenup{20}$
and the Ba{\l}uci{\'{n}}ska-Church and McCammon (1992) He cross sections, and
(b) a variable $\NHxo$ and and the Yan, Sadegpour, and Dalgarno (1998) He
cross sections.  Note the excess emission on 0.14-0.30 keV in (a) which is
absent in (b).
\label{f4}}

\clearpage
\plotone{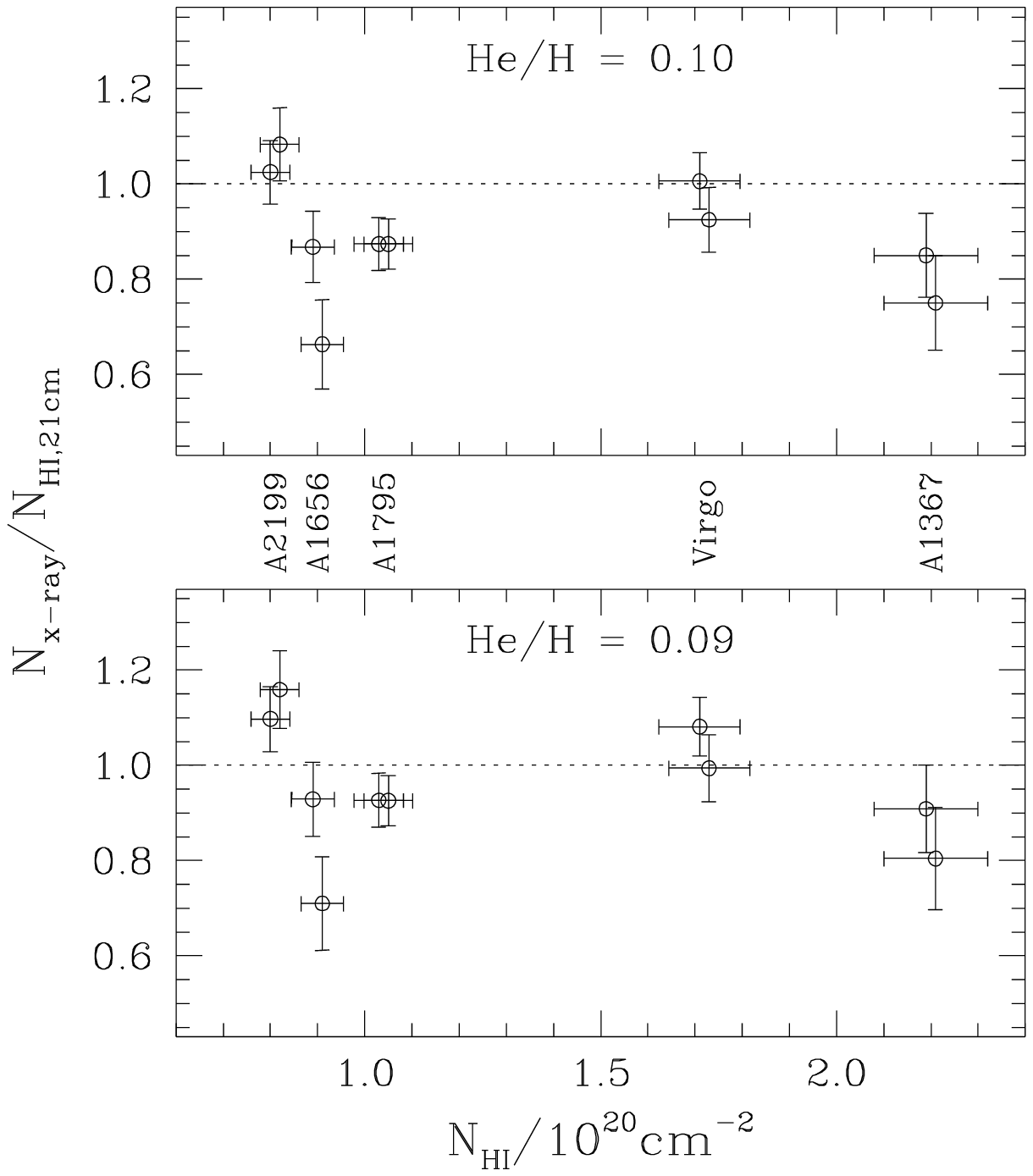}

\figcaption{$\NHxo$ toward the five clusters for two different values of
the assumed helium abundance.
\label{f5}}

\end{document}